\documentclass[%
 reprint,
 amsmath,amssymb,
 aps,
]{revtex4-1}

\usepackage{graphicx}% Include figure files
\usepackage{dcolumn}% Align table columns on decimal point
\usepackage{bm}% bold math
\usepackage{subfigure}
\begin{document}

\preprint{Arxiv}

\title{Capillary migration of microdisks on curved interfaces }% Force line breaks with \\

\author{Lu Yao}\thanks{These authors contributed equally to this work.}
\author{Nima Sharifi-Mood}\thanks{These authors contributed equally to this work.}
%\author{Lu Yao}
%\author{Nima Sharifi-Mood}
%\fntext[myfootnote]{These authors contributed equally to this work}
\author{Iris B. Liu} 
\author{Kathleen J. Stebe}
\email{kstebe@seas.upenn.edu}
%\ead{kstebe@seas.upenn.com}
\address{Chemical and Biomolecular Engineering, University of Pennsylvania, Philadelphia, PA, 19104}

\date{\today}% It is always \today, today,

\begin{abstract}
The capillary energy landscape for particles on curved fluid interfaces is strongly influenced by the particle wetting conditions. Contact line pinning has now been widely reported for colloidal particles, but its implications in capillary interactions have not been addressed. Here, we present experiment and analysis for disks with pinned contact lines on curved fluid interfaces. In experiment, we study microdisk migration on a host interface with zero mean curvature; the microdisks have contact lines pinned at their sharp edges and are sufficiently small that gravitational effects are negligible.  The disks migrate away from planar regions toward regions of steep curvature with capillary energies inferred from the dissipation along particle trajectories which are linear in the deviatoric curvature. We derive the curvature capillary energy for an interface with arbitrary curvature, and discuss each contribution to the expression. By adsorbing to a curved interface, a particle eliminates a patch of fluid interface and perturbs the surrounding interface shape. Analysis predicts that perfectly smooth,  circular disks do not migrate, and that nanometric deviations from a planar circular, contact line, like those around a weakly roughened planar disk, will drive migration with linear dependence on deviatoric curvature, in agreement with experiment.
\end{abstract}
\pacs{Valid PACS appear here}% PACS, the Physics and Astronomy
\maketitle

%\tableofcontents

\section{Introduction}

Capillary interactions are ubiquitous between microparticles at fluid interfaces. Classically, they are exploited in a wide range of technologies including stabilization of foams and Pickering emulsions \cite{Clint} and in separations of materials in ore flotation \cite{Adamson}. Fundamentally, they provide insight into condensed matter physics, for example, the formation and evolution of 2D crystal structures \cite{Pieranski,Ershov} and their topological constraints \cite{Colloidosomes}. More recently, they are exploited in bottom up assembly schemes to organize particles for advanced materials applications \cite{Tao}.\\
\indent Capillary interactions arise spontaneously between microparticles at interfaces.  Typically, the effects of gravity are negligible compared to surface tension, i.e. the particle radius $a$ is small compared to the capillary length based on the density of the fluid $\rho$, the interfacial tension $\gamma$ and the gravitational acceleration constant $g$, i.e. $\frac{a}{{\sqrt {\gamma \rho /g} }} \ll 1$. On planar interfaces, capillary interactions are well understood \cite{KralchevskyReview, Danov, Dietrich, Botto}. Particles adsorb and eliminate a patch of fluid interface.   In addition, they perturb the shape and thereby increase the area of the interface around them if they have anisotropic surface energies \cite{Lee}, non-spherical shapes \cite{Arjun,Vermant,Eric} or in principle, if they have spherical shapes but have contact lines pinned at some non-equilibrium position \cite{Stamou}. In the far field, when distortions from neighboring particles overlap, the area and therefore the energy of the interface typically decrease when the particles approach each other.\\
\indent Capillary attraction in the limit of negligible gravity has been observed to orient and assemble Janus particles \cite{Lee} and particles with complex shapes including ellipsoids \cite{Arjun,Vermant} and cylinders \cite{Eric}. Rough particles also distort planar interfaces, creating disturbances which decay close to particle contact. The energies associated with these disturbances from neighboring particles can be attractive or repulsive. Far field capillary attraction and near field capillary repulsion can be used to define equilibrium particle separation distances \cite{Lucassen,Lu}.\\
\indent There are important gaps in our understanding of curvature driven capillary migration. Fluid interfaces can be curved owing to confinement or to body forces to display mean or deviatoric curvature fields that influence particle migration.  Microparticles in the limit of negligible gravity have indeed been observed to migrate on curved interfaces. Cylindrical microparticles orient along principle axes \cite{Eric} and migrate to high curvature sites owing to their undulated contact lines \cite{Marcello}. A case of spherical microparticle migration in this limit has also been reported; the particle migrated along a curvature gradient to find an equilibrium position where capillary forces and weak gravitational forces balance \cite{Blanc}. From a theoretical perspective, the behavior of isotropic spheres with equilibrium wetting boundary conditions has been studied for several interface shapes, including interfaces with zero mean curvature \cite{Wurger}, cylindrically-shaped interfaces \cite{Dinsmore}, or interfaces with arbitrary curvatures \cite{Wurger1, Fournier, Dietrich}.  In several of these theoretical studies, particles are predicted to migrate with curvature capillary energies which are quadratic in the deviatoric curvature field; in this work, we show that contributions of that order are identically zero for the case of pinned contact lines.\\ 
\begin{figure} 
\centering
\includegraphics[width=0.5\textwidth]{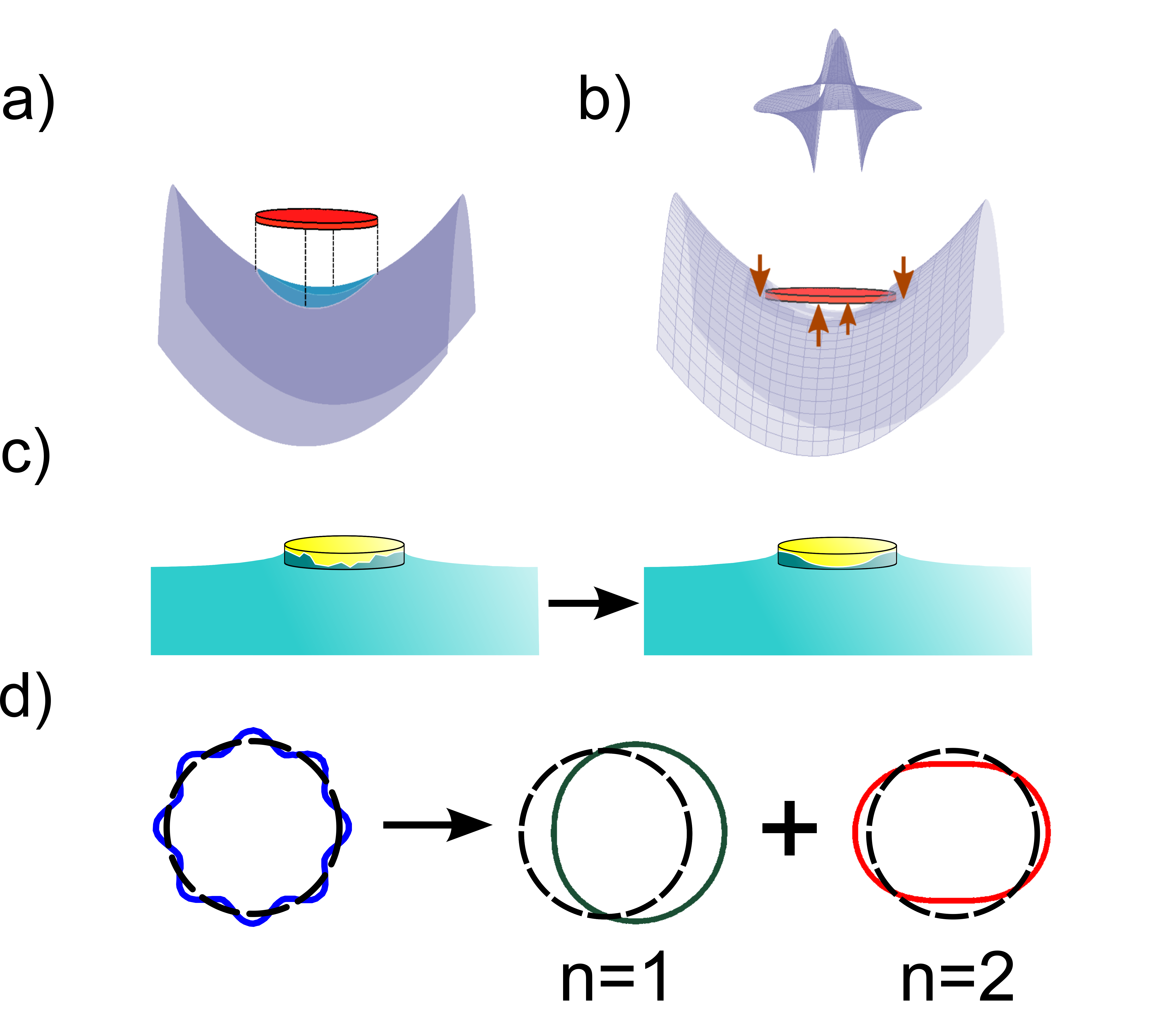}
\caption{\small{Schematic representations of (a) area of the hole under the particle in the host surface (b) curvature induced distortion of the pinned contact line (c) height undulations of contact line e.g due to particle roughness and (d) non-circular contact line (top view) shape decomposition in to dipolar (n=1, $\Gamma_1=a(1 + {\zeta _1}\cos \phi ))$, and quadrupolar (n=2, $\Gamma_2=a(1 + {\zeta _2}\cos 2\phi))$ modes.}}
 \label{SD}
\end{figure}
\indent This limit may be broadly relevant. There is a growing body of evidence that colloidal particles can have pinned contact lines at fluid interfaces in experiment \cite{Furst1, Furst2, Manoharan} and simulation \cite{Carlos, Sepideh}. Here, we study migration of particles with pinned contact lines on interfaces with well-defined curvature fields, using a planar disk-shaped particle which pins the contact line at its sharp edges. On planar interfaces, the disks interact weakly only if they are in close proximity. When placed on curved interfaces, however, these particles experience significant capillary energies and move towards regions with higher curvature. Particle trajectories are highly reproducible and deterministic for isolated disks. Particles interact in the near field to form linear structures oriented along curvature gradients.\\
\indent We derive a general theory for particles with pinned contact lines on a curved interface.  Generally, by adsorbing to a fluid interface, a particle eliminates a patch of interface, replacing it with fluid-solid surfaces. The particle also creates a disturbance in order to satisfy its pinning boundary condition at the three phase contact line.  For interfaces with finite mean curvature, this disturbance in interface height does work against the pressure jump at the interface.  For a particle of roughly circular cross section, the pinned contact line shape can be described in terms of a multipole expansion for the contact line height, with the leading order term being the quadrupolar mode \cite{Stamou} (see Fig. 1~(c) and 1~(d)).  Analysis predicts that a planar particle with a perfectly circular contact line will not migrate, and that planar particles with weak deviations from a circular contact line will migrate to regions of steepest curvature. 
 
\section{Experiments}

Epoxy resin microdisks of radius $a=5 \mu m$  are fabricated using standard lithographic methods. A negative tone photoresist (SU-8 2002, MicroChem Corp.) was deposited onto a silicon wafer (Montco Silicon) by spin-coating. The photoresist was exposed to UV light (365 nm) on a tabletop mask aligner (OAI Model 100) through a photomask (Microtronics Inc., Fine Line Imaging). The samples then were heated using UV light to  cure the polymer.  Finally, the microdisks were released from the wafer via sonication.\\
\indent Microdisks make only weak interface distortions on planar interfaces. Thus, when placed on planar interfaces, these particles fail to interact unless they are within $\sim 10$ radii of contact.  We confirm that the disks have pinned contact lines by imaging the underside of a particle trapped in an interface  (Fig.~2(a)). The image is acquired using a gel-trapping technique by placing a microparticle in the air-aqueous interface of a warm ($50^\circ C$) gellan solution which is then cooled to form a gel. PDMS is then poured over the particle, cross-linked, and lifted, making a negative mold of the interface and taking the particle with it. An SEM image of this assembly reveals that the particle pins the interface at its upper sharp edge. The disks are also rough on the nanometric scale; using an AFM (Bruker Icon) in tapping mode, we find the root mean square roughness over the surface area of the disks ranges from $18-32~nm$ (Fig.~2(b)).\\
\begin{figure} 
\includegraphics[width=.5\textwidth]{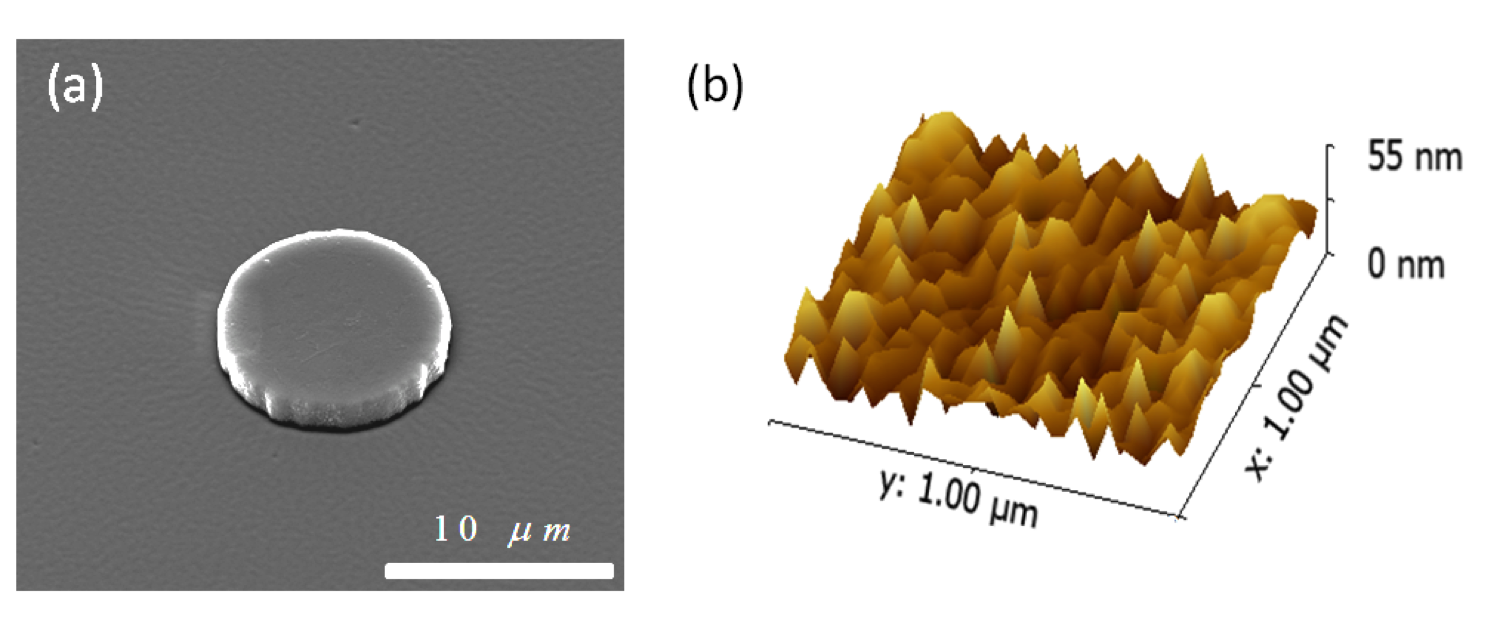}
\caption{\small{(a)  SEM image of a microdisk and PDMS negative of the air-water interface showing
contact  line  pinning. (b) AFM reveals nanoscopic roughness of the disk  surfaces of with RMS values ranging of $18-32~nm$.}}
 \label{SD1}
\end{figure}   
\indent To study particle migration, a host interface is formed with a well-defined curvature field using a technique reported previously \cite{Marcello} which we recapitulate briefly here. A curved oil-water interface is formed around a micropost of height $H_m$ and radius $R_m$ centered within a confining ring located several capillary lengths from the micropost. This structure is filled with water so that the contact line pins at the top of the micropost and the slope of the interface is $\psi \sim 15-18^{\circ}$, as measured using a goniometer (see Fig.~3(a)). The shape of the oil-water interface for distances from the micropost small compared to the capillary length obeys ${h_0} = {H_m} - {R_m}\tan \psi \ln (\frac{L}{{{R_m}}})$, where $L$ is the radial distance measured from the micropost center. This interface (shown schematically, Fig.~3(b)) has zero mean curvature ${H_0} = \frac{{{c_1} + {c_2}}}{2} = 0$, and deviatoric curvature $\Delta {c_0} = {c_1} - {c_2}$ which is finite and position dependent. The deviatoric curvature is greatest in magnitude close to the post, and decrease monotonically with distance from the post.\\
\begin{figure} 
\centering
\subfigure{\label{D1}\includegraphics[width=0.21\textwidth]{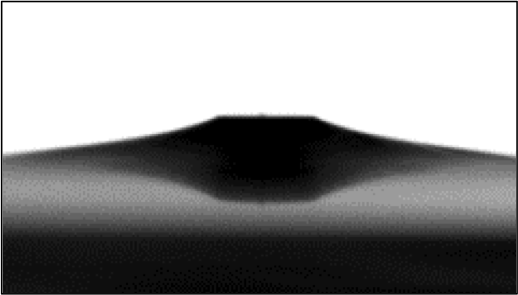}}\quad
\subfigure{\label{D2}\includegraphics[width=0.2\textwidth]{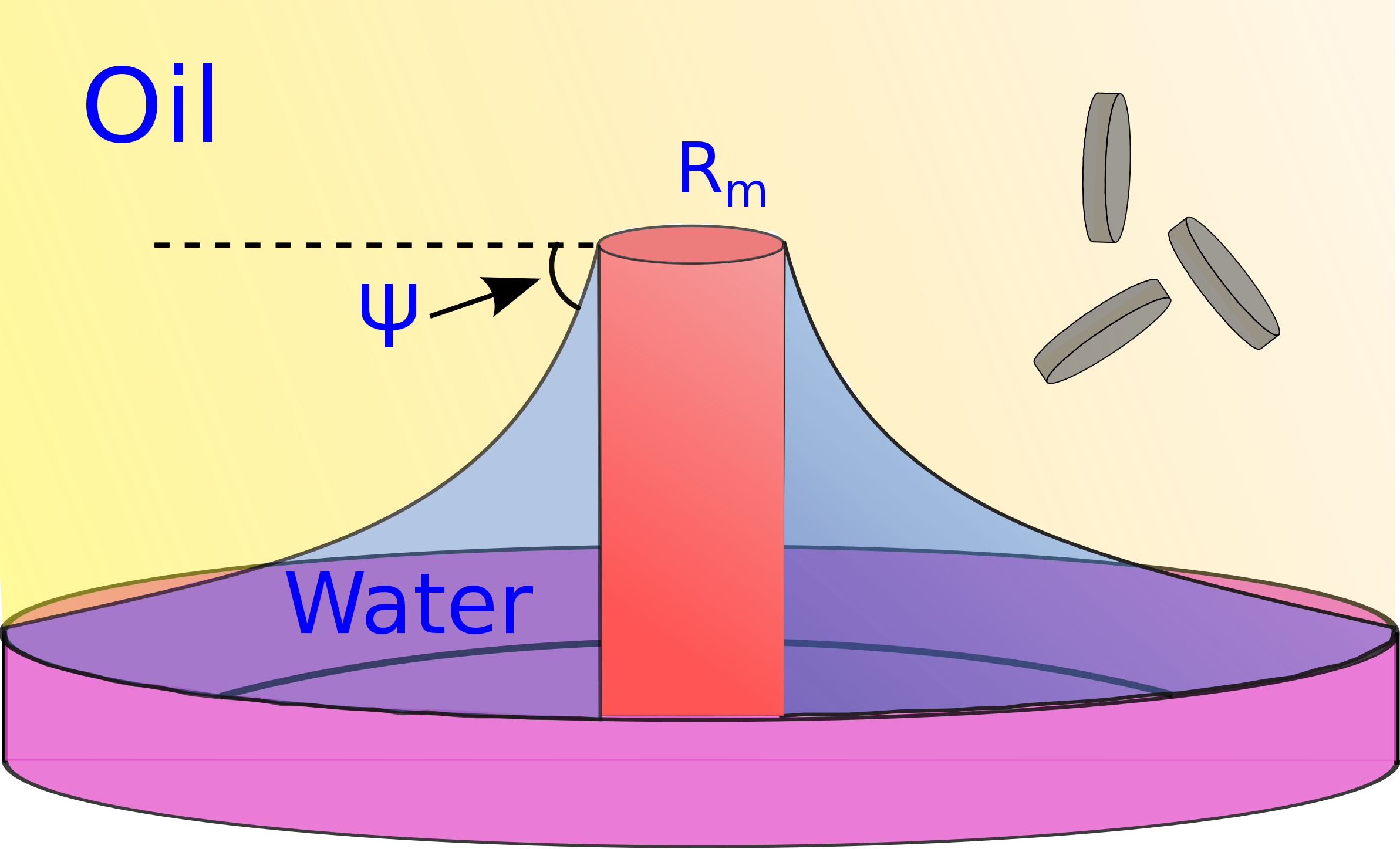}}
\caption{\small{The curved interface (a): Side view of the curved interface. (b): Schematic of the curved interface molded around a micropost. The principle curvatures are greatest in magnitude close to the post, and decrease monotonically with distance from the post.}}
 \label{SD2}
\end{figure}
\indent Disks are submerged in the oil phase, and allowed to sediment and attach to the interface. Particles, once attached, migrate up the curvature gradient along radial trajectories toward the micropost if they are isolated (Fig.~4(a)). Deviations from radial trajectories occur only for particles close to neighboring particles (within $10-15$ radii) or close to disks that are already anchored on the micropost (Fig.~4(b)).  Trajectories of isolated particles superpose when graphed in terms of $(L,t_c)$, where $t_c$ denotes time to contact defined as ${t_c} = {t_0} - t$, and $t_0$ is the time when the disk reaches the edge of the post (Fig.~4(c)). The more complex trajectories owing to interaction with neighboring elements or floating aggregates are evident as deviations from the master $L$ vs. $t_c$ curve (see Fig.~4(d)).\\
\indent Since the particle migrates in creeping flow, the capillary force on the particle is balanced by a drag force according to $F = {C_D}\mu a\frac{{dL}}{{dt}}$ where we adopt for $C_D$ Lamb's drag coefficient for a disk on an interface \cite{Lamb} and the average viscosity of the two surrounding fluids  $\mu$, is used. The capillary energy $\Delta E$ is balanced by viscous dissipation, which is found by integrating the velocity of migration over the particle position for each particle trajectory. In Fig.~5 capillary energies extracted from these trajectories for isolated disks are plotted against $a\Delta c_0$. In this figure we report only regions of the trajectory far enough from the micropost to neglect hydrodynamic interactions \cite{Brenner}. The capillary energy plotted against $a\Delta c_0$ is linear; the root mean squared error of the linear fit is approximately is $\sim 10^{-7}$ for each trajectory.  This migration is in agreement with a simple functional form which we have derived previously for migration of cylindrical microparticles migrating in curvature fields \cite{Marcello}: 
\begin{eqnarray}
E = {E_0} - \pi {a^2}\gamma (\frac{{{h_p}\Delta {c_0}}}{2}), \label{energy_capillary}
\end{eqnarray}
where $h_p$ is the amplitude of a quadrupolar distortion made by the particles. In our prior work on cylindrical microparticles, quadrupolar distortions were bounded by the equilibrium wetting configuration around the particles and thus could be associated with the particle shape. For the present case, we hypothesize that quadrupolar distortions are excited in the fluid interface owing to uncontrolled roughness of the particle, as suggested by Stamou \emph {et al.} in their analysis of particles at planar interfaces \cite{Stamou}. The data in Fig.~5 are consistent with particles with quadrupolar modes of $25-30~nm$, a magnitude similar to the roughness of the particles, suggesting that particle roughness sets the scale for disturbances sourced by the particle into the interface. The curvature capillary energy associated with this weak distortion is appreciable; $\Delta E \sim 12,000~k_BT$, explaining the non-Brownian, deterministic paths observed for the disks.\\
\indent In the experiments, the curvature of the host interface decreases monotonically with distance from the center of the micropost. For very small curvature, the capillary force driving upward migration is too small to overcome the downward pull of gravity, as reported by Blanc \emph {et al.}; if a particle attaches to an interface far away from the micropost the particle should not migrate \cite{Blanc}. We predict this threshold in our experiments by equating the gravitational force on the particle to the upward capillary force, $F_z=-\frac{\partial E}{\partial L}(\frac{dh_0}{dL})^{-1}$ and find that particles should fail to migrate for $L > 465~\mu m$ and $h_p=25~nm$. This compares well to our experimental observation ${L^{exp}}\sim308 - 449~\mu m$.\\
 \begin{figure} 
\includegraphics[width=0.48\textwidth]{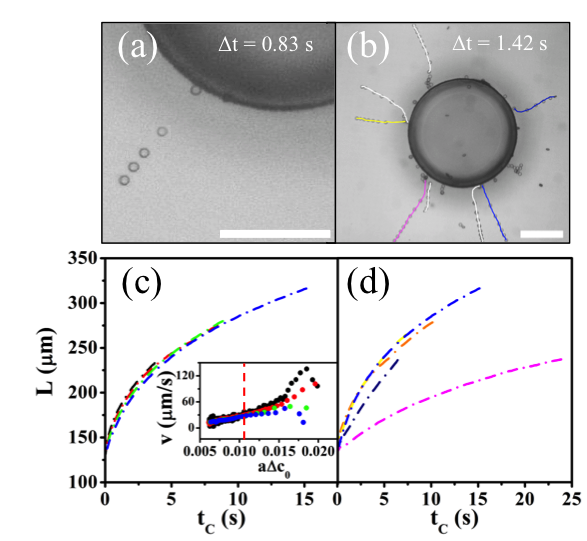}
\caption{\small{Trajectories of planar disks on the interface. Time stamped images (a) of  a microdisk trajectory (scale bar = $100\mu m$). (b) of  trajectories around a crowded micropost. Particles chain at micropost along the radial direction. (scale bar = $100\mu m$). (c) Trajectories of isolated disks migrating in the curvature field. (d) Trajectories for disks migrating on a crowded interface. Colors for trajectories in (c) and (d) correspond to trajectories of similar colors in (b).}}
 \label{SD3}
\end{figure} 
\indent Confronted by the observed linear dependence of the curvature capillary energy on $a\Delta c_0$, we derive the capillary energy in detail. Our work differs significantly from related work in the   literature for spheres which reported a quadratic dependence, i.e. proportional to ${a^2}\Delta c_0^2$. Below, we show that, while, in principle, such a term might occur, its pre-factors is identically zero. To do so, we present a detailed discussion of capillary energy for weakly non-circular pinned contact lines on curved interfaces. 
 \begin{figure} 
\includegraphics[width=0.4\textwidth]{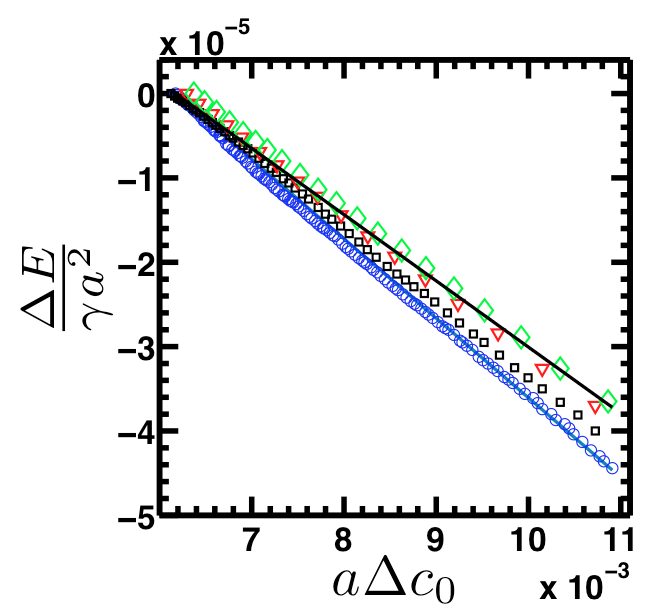}
\caption{\small{Comparison of predicted energy (Eq.~\ref{E_T_1}) (solid lines) against those extracted from experiment for trajectories in Fig.~\ref{SD3}(c) for isolated particles (symbols) (The colors correspond to the trajectories in Fig.~4(b)).  The solid black line corresponds to $h_p =25~nm$, the solid blue line corresponds to $h_p =30~nm$.}}
 \label{SD4}
\end{figure}
\section{Theory}
\subsection{Disks with weakly roughened surfaces}\label{energy_circle}
We first consider the host interface without any particle. The free energy has two contributions, 
\begin{eqnarray}
{E_1} = \gamma \mathop{{\int\!\!\!\!\!\int}\mkern-21mu \bigcirc}\limits_D 
 {(1 + \frac{{\nabla {h_0} \cdot \nabla {h_0}}}{2})dA}  - \mathop{{\int\!\!\!\!\!\int}\mkern-21mu \bigcirc}\limits_D {\Delta p~{h_0}dA},\label{E_0}
\end{eqnarray}
where $h_0$ is the height of the host interface prior to particle deposition, $\gamma$ is the interfacial tension and $D$ denotes the entire interfacial domain. In this expression, the first term is the energy due to the surface area assuming small slope $\left| {\nabla {h_0}} \right| \ll 1$, and the second term is the work done by the pressure jump across the interface $\Delta p$. When the particle is trapped at the interface, the free energy becomes 
\begin{align}
&{E_2} = {\gamma _1}{A_1} + {\gamma _2}{A_2}   \nonumber\\
&+ \gamma \mathop{{\int\!\!\!\!\!\int}\mkern-21mu \bigcirc}\limits_{D - P} 
 {(1 + \frac{{\nabla h \cdot \nabla h}}{2})dA}- \mathop{{\int\!\!\!\!\!\int}\mkern-21mu \bigcirc}\limits_D 
 {\Delta p~{h}dA} 
\end{align}
where $h$ is the height of interface after the particle is adsorbed, ${\gamma _1}{A_1}$ and ${\gamma _2}{A_2}$ are the product of the surface energies and wetted areas for the solid and the upper and lower fluids, respectively, $D$ corresponds to the domain under the particle.  The energy difference, $E$ between these two states is expressed
\begin{eqnarray}
E = {\gamma _1}{A_1} + {\gamma _2}{A_2} + \gamma [\oint\limits_{\partial (D - P)} {[\frac{1}{2}\eta \nabla \eta  + \eta \nabla {h_0}] \cdot {\bf{m}}ds} \nonumber\\
 - \gamma \mathop{{\int\!\!\!\!\!\int}\mkern-21mu \bigcirc}\limits_P 
 {(1 + \frac{{\nabla {h_0} \cdot \nabla {h_0}}}{2})dA}  + \mathop{{\int\!\!\!\!\!\int}\mkern-21mu \bigcirc}\limits_P 
 {\frac{{\Delta p}}{\gamma }({h_0} - \omega )dA}], \label{E_T}
\end{eqnarray}
where $\eta$ is the disturbance created by the particle, defined as $\eta=h-h_0$, $\omega$ is a shift of the particle center of mass perpendicular to the interface, $\partial (D - P)$ denotes the contours enclosing the domain $D-P$ and $\bf{m}$ is the unit normal to these contours tangent to the interface pointing outward the domain. In Eq.~\ref{E_T} the first term is independent of particle position, the second term is the energy owing to the disturbance $\eta$, the third term is the area of the hole in the interface created by particle adsorption and the final term is the pressure work (see Fig.~1(a) and 1(b) for detail).\\
\indent The height of the host surface at any point satisfies the Young-Laplace equation,
\begin{eqnarray}
2H\gamma  = \Delta p.\label{YL}
\end{eqnarray}
A polar coordinate $(r,\phi )$ can be defined tangent to the plane, where the origin for $\phi$ is along one of the principal axes. Adopting a Monge representation and utilizing Eq.~\ref{YL}, the interface height ${h_0}(r,\phi )$ for any surface can be expanded locally in this coordinate to be,
\begin{eqnarray}
{h_0}(r,\phi ) = \frac{{{r^2}}}{4}(2{H_0} + \Delta {c_0}\cos 2\phi ),\label{parent_exp}
\end{eqnarray}
where $H_0$ and $\Delta {c_0} = {c_1} - {c_2}$, are the mean and deviatoric curvatures of the host surface evaluated at the particle center of mass. To solve for the disturbance field, Eq.~\ref{YL} must be solved with respect to the boundary condition on the three phase contact line $\Gamma$. The leading order mode of the contact line height is a quadrupolar undulation \cite{Stamou} as the monopolar and dipolar terms are excluded by force and torque balances on the particle, respectively. This quadrupolar mode has an amplitude $h_p$ (see Fig.~1(c)), where we assume $h_p \ll a$. Higher order modes do not contribute to leading order and are not considered further.\\
\indent In the most general case the disk cross section is non-circular; deviations in the contact line from a circular shape can be described as $\Gamma  = a(1 + \sum\limits_{n = 1}^\infty  {{\zeta _n}\cos (n\phi  - {\alpha _n})} )$, where $\alpha _n$ is a phase angle and ${\zeta _n} \ll 1$. Thus, the complete boundary condition is expressed,
\begin{align}
&h(r = a[1 + \sum\limits_{n = 1}^\infty  {{\zeta _n}\cos (n\phi  - {\alpha _n})} ]) = {h_p}\frac{{{a^2}}}{{{r^2}}}\cos 2\phi,\label{bc_cl}
\end{align}
We first derive an expression for trapping energy of a disk with a perfectly circular cross-section and weak roughness, and thereafter we extend the theory for weakly non-circular cases. 
Having defined the host interface as in Eq.~\ref{parent_exp}, the area under the particle can be expressed as,
\begin{eqnarray}
\mathop{{\int\!\!\!\!\!\int}\mkern-21mu \bigcirc}\limits_P 
 {(1 + \frac{{\nabla {h_0} \cdot \nabla {h_0}}}{2})dA}  = 
 \pi {a^2} + {{\pi {a^2}}}(\frac{1}{4}{{a^2}H_0^2} + \frac{1}{16}{a^2}\Delta c_0^2),\nonumber \\ \label{hole_parent}
\end{eqnarray}
where we linearized the surface metric. The determination of the remaining terms in Eq.~\ref{E_T} requires that we obtain the disturbance $\eta$ created by the particle by solving Eq.~\ref{YL}. For circular contact lines, the pinning boundary condition can be expressed,
\begin{eqnarray}
h(r = a) = \omega  + {h_p}\cos 2\phi.
\end{eqnarray}
Far from the particle, the disturbance should decay to zero, 
\begin{eqnarray}
h(r \to \infty ) \to {h_0} = \frac{{{r^2}}}{4}\left( {2{H_0} + \Delta {c_0}\cos 2\phi } \right){\rm{ }},
\end{eqnarray}
for which the general solution is 
\begin{eqnarray}
h(r,\phi ) = \frac{{{r^2}{H_0}}}{2} + \frac{{\Delta {c_0}\cos 2\phi }}{4}({r^2} - \frac{{{a^4}}}{{{r^2}}}) + {h_p}\frac{{{a^2}}}{{{r^2}}}\cos 2\phi,\nonumber\\ \label{height_circle}
\end{eqnarray}
where 	
\begin{eqnarray}
\omega  = \frac{{{a^2}}}{2}{H_0}.
\end{eqnarray}
To confirm these analytical expressions for $h$, we have performed a numerical calculation based on a Green's function for the domain corresponding to the experimental setup, discussed in Sec.~\ref{numerical}; the numerically determined solution agrees excellently with the analysis. The disturbance due to the particle is 
\begin{eqnarray}
\eta  = {h_p}\frac{{{a^2}}}{{{r^2}}}\cos 2\phi  - \frac{{\Delta {c_0}\cos 2\phi }}{4}\frac{{{a^4}}}{{{r^2}}},\label{disturbance_circle}
\end{eqnarray}
The derivatives of $\eta$ and $h_0$ required to evaluate the remaining terms in the trapping energy are 
\begin{eqnarray}
\frac{{\partial \eta }}{{\partial r}} =  - 2{h_p}\frac{{{a^2}}}{{{r^3}}}\cos 2\phi  + \frac{{\Delta {c_0}\cos 2\phi }}{2}\frac{{{a^4}}}{{{r^3}}},
\end{eqnarray}
The contour integral given in Eq.~(\ref{E_T}) can be evaluated as, 
\begin{align}
&\oint\limits_{\partial (D - P)} {\eta \nabla \eta  \cdot {\bf{m}}ds}  = \int_0^{2\pi } {{{\left. {\frac{{\eta r}}{2}\frac{{\partial \eta }}{{\partial r}}} \right|}_{r \to \infty }}d\phi }\nonumber  \\
&- \int_0^{2\pi } {{{\left. {\frac{{\eta r}}{2}\frac{{\partial \eta }}{{\partial r}}} \right|}_{r = a}}d\phi }= 0 + \pi {a^2}[h_p^2 - \frac{{{h_p}\Delta {c_0}}}{2} + {\frac{1}{16}{{a^2\Delta c_0^2}}}],\label{self_eta}
\end{align}
The first term inside the bracket is the self-energy of the quadrupolar mode, a constant independent of curvature. The latter two terms depend on the curvature of the interface. To obtain a better insight in to the effect of deviatoric curvature, we set $H_0=0$ and note that,
\begin{align}
&-\mathop{{\int\!\!\!\!\!\int}\mkern-21mu \bigcirc}\limits_P 
 {\frac{{\nabla {h_0} \cdot \nabla {h_0}}}{2}~dA}  =- {\int_0^{2\pi } {\left. {\frac{{a{h_0}\nabla {h_0}}}{2}} \right|} _{r = a}}d\phi \nonumber \\
 &=-\frac{1}{16}\pi a^2(a^2\Delta c_0^2),\label{iden}
\end{align}
This is the excess area eliminated under the disk when it attaches to a saddle shaped interface and its magnitude is equal and opposite the quadratic term in Eq.~\ref{self_eta}. Hence, these terms cancel. We note that this is the case even for interfaces with non zero mean curvature since in calculation of Eq.~\ref{iden} terms include $H_0$ and $\Delta c_0$ never couple. The sole remaining contribution which could conceivably contribute a quadratic term in $a\Delta c_0$ is:
\begin{align}
&\oint\limits_{\partial (D - P)} {\eta \nabla {h_0} \cdot {\bf{m}}ds}  =\nonumber \\
&\int_0^{2\pi } {{{\left. {\eta r\frac{{\partial {h_0}}}{{\partial r}}} \right|}_{r \to \infty }}d\phi }  - \int_0^{2\pi } {{{\left. {a\eta \frac{{\partial {h_0}}}{{\partial r}}} \right|}_{r = a}}d\phi }\equiv 0,\label{etadh}
\end{align}
To evaluate this expression, particular care must be taken to evaluate this expression at all contours enclosing the area.  The integrand, a product of decaying and growing modes of the same power, is independent of radial position.
Therefore, the evaluation of the integral at the inner and outer contours yields a value of zero, and there are no quadratic contributions to the free energy.\\
\indent It remains to evaluate the pressure work contribution which can be written,  
\begin{align}
&\mathop{{\int\!\!\!\!\!\int}\mkern-21mu \bigcirc}\limits_P 
 {\frac{{\Delta p}}{\gamma }({h_0} - \omega )dA = 2{H_0}\int_0^a {r'dr'\int_0^{2\pi } {({h_0} - \omega )d\phi } }}\nonumber \\ 
&  = \frac{{\pi {a^4}H_0^2}}{2} - \pi {a^4}H_0^2=  - \frac{{\pi {a^4}H_0^2}}{2},\label{pressure}
\end{align}
by substituting Eq.~\ref{self_eta}, \ref{etadh} and \ref{pressure} in Eq.~\ref{E_T}, the capillary energy of a particle with pinned circular contact line can be written as, 
\begin{eqnarray}
E = {E_0} - \gamma \pi {a^2}(\frac{{3{a^2}H_0^2}}{4} + \frac{{{h_p}\Delta {c_0}}}{2}),\label{E_T_1}
\end{eqnarray}
where ${E_0} = {\gamma _1}{A_1} + {\gamma _2}{A_2} - \gamma \pi {a^2}(1 - \frac{{h_p^2}}{{{a^2}}})$, is independent of curvature.  The derivation of Eq.~\ref{E_T_1}, and its favorable comparison to experiment, is the main point of this paper.  On interfaces with constant mean curvature, Eq.~\ref{E_T_1} reduces to Eq.~\ref{energy_capillary}, which is compared to experiment in Fig. 3. Disks with pinned contact lines migrate solely because of roughness, with driving capillary energies proportional to the amplitude of the quarupolar mode of the distortion that they excite in the interface.\\
\indent Recall that the disks are not only rough but also have weakly noncircular edges. To ascertain the importance of this non-circularity, we modified the above calculation to discuss for slightly deformed disks. 
\subsection{Domain perturbation: Disks with weakly non-circular contact lines}
Returning to Eq.~\ref{E_T}, we calculate the area of the disk   which is slightly deformed. We begin by considering a contact line with a deviation from circularity which is either given by a dipolar mode $\cos\phi$ or a quadrupolar mode $\cos2\phi$. We show that the role of these distortions is weak in determining the particle migration. 
\begin{eqnarray}
{\Gamma _1} = a(1 + {\zeta _1}\cos \phi ),\label{dipole1}\\
{\Gamma _2} = a(1 + {\zeta _2}\cos 2\phi ),\label{quad1}
\end{eqnarray}
The expressions for the area of the particle corresponding to these contours are 
\begin{eqnarray}
{A_{{\Gamma _1}}} = \pi {a^2}(1 + \frac{{\zeta _1^2}}{2}),\label{A1}\\
{A_{{\Gamma _2}}} = \pi {a^2}(1 + \frac{{\zeta _2^2}}{2}),\label{A2}
\end{eqnarray}
Expressions for the area of the hole under the particle for the two contours in Eq.~\ref{dipole1} and \ref{quad1}, respectively, are 
\begin{align}
&\int_0^{2\pi } {d\phi \int_0^{a(1 + {\zeta _1}\cos \phi )} {(1 + \frac{{\nabla {h_0} \cdot \nabla {h_0}}}{2}} } )rdrd\phi \nonumber \\
&= \pi {a^2}(1 + \frac{{\zeta _1^2}}{2}) + \frac{{\pi {a^2}}}{8}(2{a^2}H_0^2 + \frac{{{a^2}\Delta c_0^2}}{2})\nonumber \\
&+ \frac{{3\pi {a^2}\zeta _1^2}}{8}(2{a^2}H_0^2 + \frac{{{a^2}\Delta c_0^2}}{2} + {a^2}{H_0}\Delta {c_0}),
\end{align}
\begin{align}
&\int_0^{2\pi } {d\phi \int_0^{a(1 + {\zeta _2}\cos 2\phi )} {(1 + \frac{{\nabla {h_0} \cdot \nabla {h_0}}}{2}} } )rdrd\phi \nonumber \\
& = \pi {a^2}(1 + \frac{{\zeta _2^2}}{2}) + \frac{{\pi {a^2}}}{8}(2{a^2}H_0^2 + \frac{{{a^2}\Delta c_0^2}}{2})\nonumber \\
 &+ \frac{{3\pi {a^2}\zeta _2^2}}{8}(2{a^2}H_0^2 + \frac{{{a^2}\Delta c_0^2}}{2}) + \frac{{\pi {a^4}{\zeta _2}{H_0}\Delta {c_0}}}{2},
\end{align}
To determine the remaining terms in Eq.~\ref{E_T}, we must evaluate $\eta$ and $\omega$; the general solution for the disturbance $\eta$ satisfying the far field boundary condition can be written as, 
\begin{align}
&\eta  = {a_0} + ({a_1}\cos \phi  + {c_1}\sin \phi )r \nonumber \\
&+ \sum\limits_{n = 1}^\infty  {({b_n}\cos n\phi  + {d_n}\sin n\phi ){r^{ - n}}},
\end{align}
and the unknown coefficient can be found by applying boundary conditions at contours given in Eq.~\ref{dipole1} and \ref{quad1} respectively. This allows evaluation of the contour integrals in Eq.~\ref{E_T}. Thus, we have, 
\begin{align}
&\oint\limits_C {(\frac{\eta }{2}\nabla \eta  + \eta \nabla {h_0}) \cdot {{\bf{m}}}ds = \pi {a^2}(\frac{{h_p^2}}{{{a^2}}} - \frac{{{h_p}\Delta c_0}}{2} + \frac{{{a^2}\Delta {c_0^2}}}{{16}})  } \nonumber\\
&+ \pi {a^2}{\zeta_1 ^2}[ - \frac{{h_p^2}}{{{a^2}}} - \frac{{5\Delta c_0{h_p}}}{4} - \frac{{H_0{h_p}}}{2} + \frac{{11H_0\Delta c_0{a^2}}}{8} \nonumber\\
&+ \frac{{5{a^2}{H_0^2}}}{2} + \frac{{7{a^2}\Delta {c_0^2}}}{8}],
\end{align}
\begin{align}
&\oint\limits_C {(\frac{\eta }{2}\nabla \eta  + \eta \nabla {h_0}) \cdot {{\bf{m}}}ds =\pi {a^2}[\frac{{h_p^2}}{{{a^2}}}- \frac{{{h_p}\Delta c_0}}{2} + \frac{{\Delta {c_0^2}{a^2}}}{{16}}]  }\nonumber\\ 
&+ \pi {a^2}\zeta_2 [\frac{{{a^2}H_0\Delta c_0}}{2} - {h_p}H_0]+ \pi {a^2}{\zeta_2 ^2}[{a^2}{H_0^2} + \frac{{7h_p^2}}{{4{a^2}}} \nonumber \\
&+ \frac{{43\Delta {c_0^2}{a^2}}}{{64}} - \frac{{13{h_p}\Delta c_0}}{8}],
\end{align}
The corresponding pressure work is:
 \begin{align}
 &\mathop{{\int\!\!\!\!\!\int}\mkern-21mu \bigcirc}\limits_P 
 {\frac{{\Delta p}}{\gamma }({h_0} - \omega )dA}  =  - \frac{{\pi {a^4}H_0^2}}{2} \nonumber\\
 &+ \frac{{\pi {a^2}\zeta _1^2}}{8}[3{a^2}{H_0}\Delta {c_0} + 8{a^2}H_0^2]
 \end{align} 
 \begin{align}
&\mathop{{\int\!\!\!\!\!\int}\mkern-21mu \bigcirc}\limits_P 
 {\frac{{\Delta p}}{\gamma }({h_0} - \omega )dA}  =  - \frac{{\pi {a^4}}H_0^2}{2} + \pi {a^4}H_0^2\zeta _2^2 +\nonumber\\
 &\frac{{3\pi {a^4}}}{2}{H_0}\Delta {c_0}{\zeta _2} - 2\pi {a^2}{H_0}{h_p}{\zeta _2},
 \end{align}
The total capillary energy for two cases can be written as, 
\begin{align}
&E = {E_0} - \pi {a^2}\gamma (\frac{{3{a^2}H_0^2}}{4} + \frac{{{h_p}\Delta c}}{2}) + \nonumber \\ 
&\pi {a^2}\zeta _1^2\gamma (\frac{{11{a^2}H_0^2}}{4} + \frac{{11{a^2}\Delta c_0^2}}{{16}} + \frac{{11{H_0}\Delta {c_0}{a^2}}}{8} \nonumber \\ 
&- \frac{{5\Delta {c_0}{h_p}}}{4} - \frac{{{H_0}{h_p}}}{2}),\label{e1}
\end{align} 
 where ${E_0} = \pi {a^2}({\gamma _1} + {\gamma _2} - \gamma )(1 + \frac{{\zeta _1^2}}{2}) + \pi h_p^2(1 - \zeta _1^2)$ and for the other case we have,
 \begin{align}
&E = {E_0} - \pi {a^2}\gamma (\frac{{3{a^2}H_0^2}}{4} + \frac{{{h_p}\Delta {c_0}}}{2}) \nonumber\\
&+ \pi {a^2}{\zeta _2}\gamma (\frac{{3{a^2}{H_0}\Delta {c_0}}}{2} - 3{H_0}{h_p})\nonumber\\
&+ \pi {a^2}\zeta _2^2\gamma (\frac{{5{a^2}H_0^2}}{4} + \frac{{31\Delta c_0^2{a^2}}}{{64}} - \frac{{13{h_p}\Delta {c_0}}}{8}),\label{e2}
\end{align}
where ${E_0} = \pi {a^2}(1 + \frac{{\zeta _2^2}}{2})({\gamma _1} + {\gamma _2} - \gamma ) + \pi h_p^2\gamma (1 + \frac{7}{4}\zeta _2^2)$, 
and from these expressions, we see that the weak perturbations in the disk radius contributes terms of $O(\zeta_1^2)$ and $O(\zeta_2)$, respectively.  These terms are negligible compared to the terms owing to particle roughness in our experiments as the magnitude of  $\zeta_1, \zeta_2 \sim 0.04$, and for interfaces with zero mean curvatures, therefore these corrections are on the order of $4-5 ~k_BT$, i.e. $\sim 10^4$ times smaller than the observed curvature capillary energies.\\ 
\indent Equations \ref{e1} and \ref{e2} are the capillary energies accounting for both roughness, which is captured by the quadrupole of magnitude $h_p$ and the leading order modes for non-circularity, captured by $\zeta_1$ and $\zeta_2$ respectively. This result was rigorously derived using singular perturbation analysis, and allows us to state the solution is strictly valid up to $O(\lambda^3)$, where $\lambda$ is the ratio of the particle size to the host interface characteristic length (typically radius of curvature).
\subsection{Numerical analysis based on singularity solution}\label{numerical}
 \begin{figure} 
 \center
\includegraphics[width=0.47\textwidth]{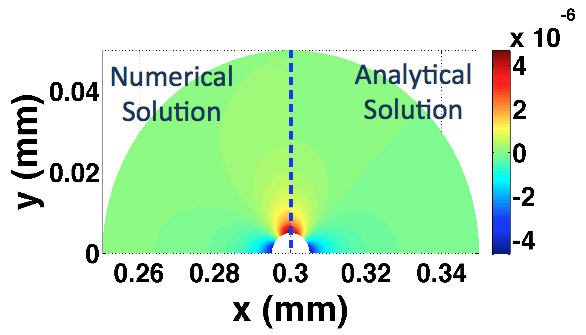}
\caption{Disturbance $\eta =h-h_0$ around a circular disk on a curved host surface (similar to our experiment) with zero mean curvature found from numerical analysis declared in Sec. \ref{numerical} (left side) and analytical solution for $h_p = 0$ given in Sec. \ref{energy_circle} (right side).}
 \label{Co1}
\end{figure}
We have numerically confirmed the form for the disturbance $\eta$ found in Sec.~\ref{energy_circle} for a disk with perfect circular cross section on a curved interface with geometry that corresponds to our experiments ${h_0} = {H_m} - {R_m}\tan \psi \ln (\frac{L}{{{R_m}}})$ where $L$ is the radial distance from the center of micropost. We numerically investigate the disturbance field created by a small circular disk with pinned contact line placed on a curved interface.  The interface geometry was fixed, as in our experiments, by pinning on a tall cylindrical micropost and an outer ring (see schematic figure in Fig.~3b). For small slopes, the height of interface at any point can be governed via the following integral equation, 
\begin{align}
&h({\bf{r}}) = \int {\int {G({\bf{r}},{\bf{r'}})\rho ({\bf{r'}})r'dr'd\phi '} }\nonumber\\
&\mp {R_i}\int {{{\left. {h({\bf{r'}})\frac{{\partial G({\bf{r}},{\bf{r'}})}}{{\partial r'}}} \right|}_{r' = {R_i}}}d\phi '},\label{boundaryintegral}
\end{align}  
where $G({\bf{r}},{\bf{r'}})$ is the Green's function of $\nabla^2$ operator with Dirichlet boundary conditions, which by definition satisfies,
\begin{eqnarray}
{\nabla ^2}G({\bf{r}},{\bf{r'}}) =  - {\delta ^2}({\bf{r}} - {\bf{r'}}),
\end{eqnarray}
where ${\delta ^2}({\bf{r}} - {\bf{r'}})$ is a two dimensional Dirac delta function. We first developed the Green's function with homogeneous Dirichlet boundary conditions at the micropost and outer ring, and thereafter we introduce the boundary condition at the disk with $N$ capillary charge singularities located at its circumference as, 
\begin{eqnarray}
\rho ({\bf{r}}) = \sum\limits_{m = 1}^N {{Q_m}\frac{{\delta (r - {r_{{q_m}}})\delta (\phi  - {\phi _{{q_m}}})}}{r}},
\end{eqnarray}
where the magnitude of each charge, $Q_m$, is unknown and should be determined via Eq.~\ref{boundaryintegral} by imposing the pinning boundary condition. We then solve the set of algebraic linear equations for $Q_m$ to find the charge magnitudes. Therefore, we are able to obtain the height at any point in domain via Eq.~\ref{boundaryintegral}. Our numerical solution for the disturbance created by the particle in the interface agrees with the analytical solution, to with 2\%, as shown in Fig.~6.  This confirms the analytical form for the disturbance created by the particle used to calculate the curvature capillary energy.\\
\section{Conclusions}

We have performed experiments using disks with pinned contact lines on an interface with strong deviatoric curvature gradients. The particles migrated up to sites of high curvature with capillary energies that are linear in the deviatoric curvature $\Delta c_0$.  This experimental finding is in contrast to prior work on spheres with equilibrium contact lines in the interface.  To place this result in context, we derived an analytical expression for the curvature capillary migration energy, and show that it is indeed linear in the product $h_p \Delta c_0$, where $h_p$ is the amplitude of the quadrupolar mode of  the particle sourced disturbance.  The values for $h_p$ inferred from the energy data correspond to the scale of the particle roughness as determined by AFM.  Our analysis shows definitively that no quadratic term in deviatoric curvature is present in the migration energy to leading order, a marked difference from prior work in the literature. This analysis can be extended to pair interactions of particles on curved interfaces.  Qualitatively, the formation of chains along the curvature gradient direction evident in Fig.~4(b) can readily be explained.  The quadrupole owing to roughness aligns along the principle axes of the interface \cite{Eric}, so interacting particles chain along the radial direction. Finally, these results for disks have important implications for spheres with pinned contact lines; we address this issue in a separate communication.

This work is partially supported by NSF grants CBET-1066284 and CBET-1133267 and GAANN P200A120246.

\section*{References}

\begin{thebibliography}{31}%
\makeatletter
\providecommand \@ifxundefined [1]{%
 \@ifx{#1\undefined}
}%
\providecommand \@ifnum [1]{%
 \ifnum #1\expandafter \@firstoftwo
 \else \expandafter \@secondoftwo
 \fi
}%
\providecommand \@ifx [1]{%
 \ifx #1\expandafter \@firstoftwo
 \else \expandafter \@secondoftwo
 \fi
}%
\providecommand \natexlab [1]{#1}%
\providecommand \enquote  [1]{``#1''}%
\providecommand \bibnamefont  [1]{#1}%
\providecommand \bibfnamefont [1]{#1}%
\providecommand \citenamefont [1]{#1}%
\providecommand \href@noop [0]{\@secondoftwo}%
\providecommand \href [0]{\begingroup \@sanitize@url \@href}%
\providecommand \@href[1]{\@@startlink{#1}\@@href}%
\providecommand \@@href[1]{\endgroup#1\@@endlink}%
\providecommand \@sanitize@url [0]{\catcode `\\12\catcode `\$12\catcode
  `\&12\catcode `\#12\catcode `\^12\catcode `\_12\catcode `\%12\relax}%
\providecommand \@@startlink[1]{}%
\providecommand \@@endlink[0]{}%
\providecommand \url  [0]{\begingroup\@sanitize@url \@url }%
\providecommand \@url [1]{\endgroup\@href {#1}{\urlprefix }}%
\providecommand \urlprefix  [0]{URL }%
\providecommand \Eprint [0]{\href }%
\providecommand \doibase [0]{http://dx.doi.org/}%
\providecommand \selectlanguage [0]{\@gobble}%
\providecommand \bibinfo  [0]{\@secondoftwo}%
\providecommand \bibfield  [0]{\@secondoftwo}%
\providecommand \translation [1]{[#1]}%
\providecommand \BibitemOpen [0]{}%
\providecommand \bibitemStop [0]{}%
\providecommand \bibitemNoStop [0]{.\EOS\space}%
\providecommand \EOS [0]{\spacefactor3000\relax}%
\providecommand \BibitemShut  [1]{\csname bibitem#1\endcsname}%
\let\auto@bib@innerbib\@empty
%</preamble>
\bibitem [{\citenamefont {Aveyard}\ \emph {et~al.}(2003)\citenamefont
  {Aveyard}, \citenamefont {Binks},\ and\ \citenamefont {Clint}}]{Clint}%
  \BibitemOpen
  \bibfield  {author} {\bibinfo {author} {\bibfnamefont {R.}~\bibnamefont
  {Aveyard}}, \bibinfo {author} {\bibfnamefont {B.~P.}\ \bibnamefont {Binks}},
  \ and\ \bibinfo {author} {\bibfnamefont {J.~H.}\ \bibnamefont {Clint}},\
  }\href {\doibase http://dx.doi.org/10.1016/S0001-8686(02)00069-6} {\bibfield
  {journal} {\bibinfo  {journal} {Adv. Colloid Interface Sci.}\ }\textbf
  {\bibinfo {volume} {100}},\ \bibinfo {pages} {503} (\bibinfo {year}
  {2003})}\BibitemShut {NoStop}%
\bibitem [{\citenamefont {Adamson}\ and\ \citenamefont {Gast}(1997)}]{Adamson}%
  \BibitemOpen
  \bibfield  {author} {\bibinfo {author} {\bibfnamefont {A.~W.}\ \bibnamefont
  {Adamson}}\ and\ \bibinfo {author} {\bibfnamefont {A.~P.}\ \bibnamefont
  {Gast}},\ }\href@noop {} {\emph {\bibinfo {title} {Physical Chemistry of
  Surfaces}}}\ (\bibinfo  {publisher} {Wiley},\ \bibinfo {address} {New York,
  NY, USA},\ \bibinfo {year} {1997})\BibitemShut {NoStop}%
\bibitem [{\citenamefont {Pieranski}(1980)}]{Pieranski}%
  \BibitemOpen
  \bibfield  {author} {\bibinfo {author} {\bibfnamefont {P.}~\bibnamefont
  {Pieranski}},\ }\href {\doibase 10.1103/PhysRevLett.45.569} {\bibfield
  {journal} {\bibinfo  {journal} {Phys. Rev. Lett.}\ }\textbf {\bibinfo
  {volume} {45}},\ \bibinfo {pages} {569} (\bibinfo {year} {1980})}\BibitemShut
  {NoStop}%
\bibitem [{\citenamefont {Ershov}\ \emph {et~al.}(2013)\citenamefont {Ershov},
  \citenamefont {Sprakel}, \citenamefont {Appel}, \citenamefont
  {Cohen~Stuart},\ and\ \citenamefont {van~der Gucht}}]{Ershov}%
  \BibitemOpen
  \bibfield  {author} {\bibinfo {author} {\bibfnamefont {D.}~\bibnamefont
  {Ershov}}, \bibinfo {author} {\bibfnamefont {J.}~\bibnamefont {Sprakel}},
  \bibinfo {author} {\bibfnamefont {J.}~\bibnamefont {Appel}}, \bibinfo
  {author} {\bibfnamefont {M.~A.}\ \bibnamefont {Cohen~Stuart}}, \ and\
  \bibinfo {author} {\bibfnamefont {J.}~\bibnamefont {van~der Gucht}},\ }\href
  {\doibase 10.1073/pnas.1222196110} {\bibfield  {journal} {\bibinfo  {journal}
  {Proc. Natl. Acad. Sci.}\ }\textbf {\bibinfo {volume} {110}},\ \bibinfo
  {pages} {9220} (\bibinfo {year} {2013})}\BibitemShut {NoStop}%
\bibitem [{\citenamefont {Dinsmore}\ \emph {et~al.}(2002)\citenamefont
  {Dinsmore}, \citenamefont {Hsu}, \citenamefont {Nikolaides}, \citenamefont
  {Marquez}, \citenamefont {Bausch},\ and\ \citenamefont
  {Weitz}}]{Colloidosomes}%
  \BibitemOpen
  \bibfield  {author} {\bibinfo {author} {\bibfnamefont {A.~D.}\ \bibnamefont
  {Dinsmore}}, \bibinfo {author} {\bibfnamefont {M.~F.}\ \bibnamefont {Hsu}},
  \bibinfo {author} {\bibfnamefont {M.~G.}\ \bibnamefont {Nikolaides}},
  \bibinfo {author} {\bibfnamefont {M.}~\bibnamefont {Marquez}}, \bibinfo
  {author} {\bibfnamefont {A.~R.}\ \bibnamefont {Bausch}}, \ and\ \bibinfo
  {author} {\bibfnamefont {D.~A.}\ \bibnamefont {Weitz}},\ }\href {\doibase
  10.1126/science.1074868} {\bibfield  {journal} {\bibinfo  {journal}
  {Science}\ }\textbf {\bibinfo {volume} {298}},\ \bibinfo {pages} {1006}
  (\bibinfo {year} {2002})}\BibitemShut {NoStop}%
\bibitem [{\citenamefont {Tao}\ \emph {et~al.}(2007)\citenamefont {Tao},
  \citenamefont {Sinsermsuksakul},\ and\ \citenamefont {Yang}}]{Tao}%
  \BibitemOpen
  \bibfield  {author} {\bibinfo {author} {\bibfnamefont {A.}~\bibnamefont
  {Tao}}, \bibinfo {author} {\bibfnamefont {P.}~\bibnamefont
  {Sinsermsuksakul}}, \ and\ \bibinfo {author} {\bibfnamefont {P.}~\bibnamefont
  {Yang}},\ }\href {http://dx.doi.org/10.1038/nnano.2007.189} {\bibfield
  {journal} {\bibinfo  {journal} {Nat Nano}\ }\textbf {\bibinfo {volume} {2}},\
  \bibinfo {pages} {435} (\bibinfo {year} {2007})}\BibitemShut {NoStop}%
\bibitem [{\citenamefont {Kralchevsky}\ and\ \citenamefont
  {Nagayama}(2000)}]{KralchevskyReview}%
  \BibitemOpen
  \bibfield  {author} {\bibinfo {author} {\bibfnamefont {P.~A.}\ \bibnamefont
  {Kralchevsky}}\ and\ \bibinfo {author} {\bibfnamefont {K.}~\bibnamefont
  {Nagayama}},\ }\href {\doibase
  http://dx.doi.org/10.1016/S0001-8686(99)00016-0} {\bibfield  {journal}
  {\bibinfo  {journal} {Adv. Colloid Interface Sci.}\ }\textbf {\bibinfo
  {volume} {85}},\ \bibinfo {pages} {145} (\bibinfo {year} {2000})}\BibitemShut
  {NoStop}%
\bibitem [{\citenamefont {Danov}\ and\ \citenamefont
  {Kralchevsky}(2010)}]{Danov}%
  \BibitemOpen
  \bibfield  {author} {\bibinfo {author} {\bibfnamefont {K.~D.}\ \bibnamefont
  {Danov}}\ and\ \bibinfo {author} {\bibfnamefont {P.~A.~A.}\ \bibnamefont
  {Kralchevsky}},\ }\href {\doibase
  http://dx.doi.org/10.1016/j.cis.2010.01.010} {\bibfield  {journal} {\bibinfo
  {journal} {Adv. Colloid Interface Sci.}\ }\textbf {\bibinfo {volume} {154}},\
  \bibinfo {pages} {91} (\bibinfo {year} {2010})}\BibitemShut {NoStop}%
\bibitem [{\citenamefont {Dom{\'\i}nguez}\ \emph {et~al.}(2008)\citenamefont
  {Dom{\'\i}nguez}, \citenamefont {Oettel},\ and\ \citenamefont
  {Dietrich}}]{Dietrich}%
  \BibitemOpen
  \bibfield  {author} {\bibinfo {author} {\bibfnamefont {A.}~\bibnamefont
  {Dom{\'\i}nguez}}, \bibinfo {author} {\bibfnamefont {M.}~\bibnamefont
  {Oettel}}, \ and\ \bibinfo {author} {\bibfnamefont {S.}~\bibnamefont
  {Dietrich}},\ }\href {\doibase http://dx.doi.org/10.1063/1.2890035}
  {\bibfield  {journal} {\bibinfo  {journal} {J. Chem. Phys.}\ }\textbf
  {\bibinfo {volume} {128}},\ \bibinfo {pages} {114904} (\bibinfo {year}
  {2008})}\BibitemShut {NoStop}%
\bibitem [{\citenamefont {Botto}\ \emph {et~al.}(2012)\citenamefont {Botto},
  \citenamefont {Lewandowski}, \citenamefont {Cavallaro},\ and\ \citenamefont
  {Stebe}}]{Botto}%
  \BibitemOpen
  \bibfield  {author} {\bibinfo {author} {\bibfnamefont {L.}~\bibnamefont
  {Botto}}, \bibinfo {author} {\bibfnamefont {E.~P.}\ \bibnamefont
  {Lewandowski}}, \bibinfo {author} {\bibfnamefont {M.}~\bibnamefont
  {Cavallaro}}, \ and\ \bibinfo {author} {\bibfnamefont {K.~J.}\ \bibnamefont
  {Stebe}},\ }\href {\doibase 10.1039/C2SM25929J} {\bibfield  {journal}
  {\bibinfo  {journal} {Soft Matter}\ }\textbf {\bibinfo {volume} {8}},\
  \bibinfo {pages} {9957} (\bibinfo {year} {2012})}\BibitemShut {NoStop}%
\bibitem [{\citenamefont {Park}\ \emph {et~al.}(2011)\citenamefont {Park},
  \citenamefont {Brugarolas},\ and\ \citenamefont {Lee}}]{Lee}%
  \BibitemOpen
  \bibfield  {author} {\bibinfo {author} {\bibfnamefont {B.}~\bibnamefont
  {Park}}, \bibinfo {author} {\bibfnamefont {T.}~\bibnamefont {Brugarolas}}, \
  and\ \bibinfo {author} {\bibfnamefont {D.}~\bibnamefont {Lee}},\ }\href
  {\doibase 10.1039/C1SM05460K} {\bibfield  {journal} {\bibinfo  {journal}
  {Soft Matter}\ }\textbf {\bibinfo {volume} {7}},\ \bibinfo {pages} {6413}
  (\bibinfo {year} {2011})}\BibitemShut {NoStop}%
\bibitem [{\citenamefont {Loudet}\ \emph {et~al.}(2005)\citenamefont {Loudet},
  \citenamefont {Alsayed}, \citenamefont {Zhang},\ and\ \citenamefont
  {Yodh}}]{Arjun}%
  \BibitemOpen
  \bibfield  {author} {\bibinfo {author} {\bibfnamefont {J.~C.}\ \bibnamefont
  {Loudet}}, \bibinfo {author} {\bibfnamefont {A.~M.}\ \bibnamefont {Alsayed}},
  \bibinfo {author} {\bibfnamefont {J.}~\bibnamefont {Zhang}}, \ and\ \bibinfo
  {author} {\bibfnamefont {A.~G.}\ \bibnamefont {Yodh}},\ }\href {\doibase
  10.1103/PhysRevLett.94.018301} {\bibfield  {journal} {\bibinfo  {journal}
  {Phys. Rev. Lett.}\ }\textbf {\bibinfo {volume} {94}},\ \bibinfo {pages}
  {018301} (\bibinfo {year} {2005})}\BibitemShut {NoStop}%
\bibitem [{\citenamefont {Zhang}\ \emph {et~al.}(2011)\citenamefont {Zhang},
  \citenamefont {Pfleiderer}, \citenamefont {Schofield}, \citenamefont
  {Clasen},\ and\ \citenamefont {Vermant}}]{Vermant}%
  \BibitemOpen
  \bibfield  {author} {\bibinfo {author} {\bibfnamefont {Z.}~\bibnamefont
  {Zhang}}, \bibinfo {author} {\bibfnamefont {P.}~\bibnamefont {Pfleiderer}},
  \bibinfo {author} {\bibfnamefont {A.~B.}\ \bibnamefont {Schofield}}, \bibinfo
  {author} {\bibfnamefont {C.}~\bibnamefont {Clasen}}, \ and\ \bibinfo {author}
  {\bibfnamefont {J.}~\bibnamefont {Vermant}},\ }\href {\doibase
  10.1021/ja108099r} {\bibfield  {journal} {\bibinfo  {journal} {J. Am. Chem.
  Soc.}\ }\textbf {\bibinfo {volume} {133}},\ \bibinfo {pages} {392} (\bibinfo
  {year} {2011})}\BibitemShut {NoStop}%
\bibitem [{\citenamefont {Lewandowski}\ \emph {et~al.}(2008)\citenamefont
  {Lewandowski}, \citenamefont {Bernate}, \citenamefont {Searson},\ and\
  \citenamefont {Stebe}}]{Eric}%
  \BibitemOpen
  \bibfield  {author} {\bibinfo {author} {\bibfnamefont {E.~P.}\ \bibnamefont
  {Lewandowski}}, \bibinfo {author} {\bibfnamefont {J.~A.}\ \bibnamefont
  {Bernate}}, \bibinfo {author} {\bibfnamefont {P.~C.}\ \bibnamefont
  {Searson}}, \ and\ \bibinfo {author} {\bibfnamefont {K.~J.}\ \bibnamefont
  {Stebe}},\ }\href {\doibase 10.1021/la801167h} {\bibfield  {journal}
  {\bibinfo  {journal} {Langmuir}\ }\textbf {\bibinfo {volume} {24}},\ \bibinfo
  {pages} {9302} (\bibinfo {year} {2008})}\BibitemShut {NoStop}%
\bibitem [{\citenamefont {Stamou}\ \emph {et~al.}(2000)\citenamefont {Stamou},
  \citenamefont {Duschl},\ and\ \citenamefont {Johannsmann}}]{Stamou}%
  \BibitemOpen
  \bibfield  {author} {\bibinfo {author} {\bibfnamefont {D.}~\bibnamefont
  {Stamou}}, \bibinfo {author} {\bibfnamefont {C.}~\bibnamefont {Duschl}}, \
  and\ \bibinfo {author} {\bibfnamefont {D.}~\bibnamefont {Johannsmann}},\
  }\href {\doibase 10.1103/PhysRevE.62.5263} {\bibfield  {journal} {\bibinfo
  {journal} {Phys. Rev. E}\ }\textbf {\bibinfo {volume} {62}},\ \bibinfo
  {pages} {5263} (\bibinfo {year} {2000})}\BibitemShut {NoStop}%
\bibitem [{\citenamefont {Lucassen}(1992)}]{Lucassen}%
  \BibitemOpen
  \bibfield  {author} {\bibinfo {author} {\bibfnamefont {J.}~\bibnamefont
  {Lucassen}},\ }\href {\doibase
  http://dx.doi.org/10.1016/0166-6622(92)80268-7} {\bibfield  {journal}
  {\bibinfo  {journal} {Colloids and Surfaces}\ }\textbf {\bibinfo {volume}
  {65}},\ \bibinfo {pages} {131} (\bibinfo {year} {1992})}\BibitemShut
  {NoStop}%
\bibitem [{\citenamefont {Yao}\ \emph {et~al.}(2013)\citenamefont {Yao},
  \citenamefont {Botto}, \citenamefont {Cavallaro}, \citenamefont {Bleier},
  \citenamefont {Garbin},\ and\ \citenamefont {Stebe}}]{Lu}%
  \BibitemOpen
  \bibfield  {author} {\bibinfo {author} {\bibfnamefont {L.}~\bibnamefont
  {Yao}}, \bibinfo {author} {\bibfnamefont {L.}~\bibnamefont {Botto}}, \bibinfo
  {author} {\bibfnamefont {M.}~\bibnamefont {Cavallaro}}, \bibinfo {author}
  {\bibfnamefont {B.~J.}\ \bibnamefont {Bleier}}, \bibinfo {author}
  {\bibfnamefont {V.}~\bibnamefont {Garbin}}, \ and\ \bibinfo {author}
  {\bibfnamefont {K.~J.}\ \bibnamefont {Stebe}},\ }\href {\doibase
  10.1039/C2SM27020J} {\bibfield  {journal} {\bibinfo  {journal} {Soft Matter}\
  }\textbf {\bibinfo {volume} {9}},\ \bibinfo {pages} {779} (\bibinfo {year}
  {2013})}\BibitemShut {NoStop}%
\bibitem [{\citenamefont {Cavallaro}\ \emph {et~al.}(2011)\citenamefont
  {Cavallaro}, \citenamefont {Botto}, \citenamefont {Lewandowski},
  \citenamefont {Wang},\ and\ \citenamefont {Stebe}}]{Marcello}%
  \BibitemOpen
  \bibfield  {author} {\bibinfo {author} {\bibfnamefont {M.}~\bibnamefont
  {Cavallaro}}, \bibinfo {author} {\bibfnamefont {L.}~\bibnamefont {Botto}},
  \bibinfo {author} {\bibfnamefont {E.~P.}\ \bibnamefont {Lewandowski}},
  \bibinfo {author} {\bibfnamefont {M.}~\bibnamefont {Wang}}, \ and\ \bibinfo
  {author} {\bibfnamefont {K.~J.}\ \bibnamefont {Stebe}},\ }\href {\doibase
  10.1073/pnas.1116344108} {\bibfield  {journal} {\bibinfo  {journal} {Proc.
  Natl. Acad. Sci.}\ }\textbf {\bibinfo {volume} {108}},\ \bibinfo {pages}
  {20923} (\bibinfo {year} {2011})}\BibitemShut {NoStop}%
\bibitem [{\citenamefont {Blanc}\ \emph {et~al.}(2013)\citenamefont {Blanc},
  \citenamefont {Fedorenko}, \citenamefont {Gross}, \citenamefont {In},
  \citenamefont {Abkarian}, \citenamefont {Gharbi}, \citenamefont {Fournier},
  \citenamefont {Galatola},\ and\ \citenamefont {Nobili}}]{Blanc}%
  \BibitemOpen
  \bibfield  {author} {\bibinfo {author} {\bibfnamefont {C.}~\bibnamefont
  {Blanc}}, \bibinfo {author} {\bibfnamefont {D.}~\bibnamefont {Fedorenko}},
  \bibinfo {author} {\bibfnamefont {M.}~\bibnamefont {Gross}}, \bibinfo
  {author} {\bibfnamefont {M.}~\bibnamefont {In}}, \bibinfo {author}
  {\bibfnamefont {M.}~\bibnamefont {Abkarian}}, \bibinfo {author}
  {\bibfnamefont {M.}~\bibnamefont {Gharbi}}, \bibinfo {author} {\bibfnamefont
  {J.}~\bibnamefont {Fournier}}, \bibinfo {author} {\bibfnamefont
  {P.}~\bibnamefont {Galatola}}, \ and\ \bibinfo {author} {\bibfnamefont
  {M.}~\bibnamefont {Nobili}},\ }\href {\doibase
  10.1103/PhysRevLett.111.058302} {\bibfield  {journal} {\bibinfo  {journal}
  {Phys. Rev. Lett.}\ }\textbf {\bibinfo {volume} {111}},\ \bibinfo {pages}
  {058302} (\bibinfo {year} {2013})}\BibitemShut {NoStop}%
\bibitem [{\citenamefont {W\"urger}(2006)}]{Wurger}%
  \BibitemOpen
  \bibfield  {author} {\bibinfo {author} {\bibfnamefont {A.}~\bibnamefont
  {W\"urger}},\ }\href {\doibase 10.1103/PhysRevE.74.041402} {\bibfield
  {journal} {\bibinfo  {journal} {Phys. Rev. E}\ }\textbf {\bibinfo {volume}
  {74}},\ \bibinfo {pages} {041402} (\bibinfo {year} {2006})}\BibitemShut
  {NoStop}%
\bibitem [{\citenamefont {Zeng}\ \emph {et~al.}(2012)\citenamefont {Zeng},
  \citenamefont {Brau}, \citenamefont {Davidovitch},\ and\ \citenamefont
  {Dinsmore}}]{Dinsmore}%
  \BibitemOpen
  \bibfield  {author} {\bibinfo {author} {\bibfnamefont {C.}~\bibnamefont
  {Zeng}}, \bibinfo {author} {\bibfnamefont {F.}~\bibnamefont {Brau}}, \bibinfo
  {author} {\bibfnamefont {B.}~\bibnamefont {Davidovitch}}, \ and\ \bibinfo
  {author} {\bibfnamefont {A.~D.}\ \bibnamefont {Dinsmore}},\ }\href {\doibase
  10.1039/C2SM25871D} {\bibfield  {journal} {\bibinfo  {journal} {Soft Matter}\
  }\textbf {\bibinfo {volume} {8}},\ \bibinfo {pages} {8582} (\bibinfo {year}
  {2012})}\BibitemShut {NoStop}%
\bibitem [{\citenamefont {L{\'e}andri}\ and\ \citenamefont
  {W{\"u}rger}(2013)}]{Wurger1}%
  \BibitemOpen
  \bibfield  {author} {\bibinfo {author} {\bibfnamefont {J.}~\bibnamefont
  {L{\'e}andri}}\ and\ \bibinfo {author} {\bibfnamefont {A.}~\bibnamefont
  {W{\"u}rger}},\ }\href {\doibase
  http://dx.doi.org/10.1016/j.jcis.2013.04.024} {\bibfield  {journal} {\bibinfo
   {journal} {J. Colloid Interface Sci.}\ }\textbf {\bibinfo {volume} {405}},\
  \bibinfo {pages} {249} (\bibinfo {year} {2013})}\BibitemShut {NoStop}%
\bibitem [{\citenamefont {Galatola}\ and\ \citenamefont
  {Fournier}(2014)}]{Fournier}%
  \BibitemOpen
  \bibfield  {author} {\bibinfo {author} {\bibfnamefont {P.}~\bibnamefont
  {Galatola}}\ and\ \bibinfo {author} {\bibfnamefont {J.}~\bibnamefont
  {Fournier}},\ }\href {\doibase 10.1039/C3SM52622D} {\bibfield  {journal}
  {\bibinfo  {journal} {Soft Matter}\ }\textbf {\bibinfo {volume} {10}},\
  \bibinfo {pages} {2197} (\bibinfo {year} {2014})}\BibitemShut {NoStop}%
\bibitem [{\citenamefont {Park}\ and\ \citenamefont {Furst}(2011)}]{Furst1}%
  \BibitemOpen
  \bibfield  {author} {\bibinfo {author} {\bibfnamefont {B.}~\bibnamefont
  {Park}}\ and\ \bibinfo {author} {\bibfnamefont {E.~M.}\ \bibnamefont
  {Furst}},\ }\href {\doibase 10.1039/C1SM00005E} {\bibfield  {journal}
  {\bibinfo  {journal} {Soft Matter}\ }\textbf {\bibinfo {volume} {7}},\
  \bibinfo {pages} {7676} (\bibinfo {year} {2011})}\BibitemShut {NoStop}%
\bibitem [{\citenamefont {Park}\ \emph {et~al.}(2010)\citenamefont {Park},
  \citenamefont {Vermant},\ and\ \citenamefont {Furst}}]{Furst2}%
  \BibitemOpen
  \bibfield  {author} {\bibinfo {author} {\bibfnamefont {B.}~\bibnamefont
  {Park}}, \bibinfo {author} {\bibfnamefont {J.}~\bibnamefont {Vermant}}, \
  and\ \bibinfo {author} {\bibfnamefont {E.~M.}\ \bibnamefont {Furst}},\ }\href
  {\doibase 10.1039/C0SM00485E} {\bibfield  {journal} {\bibinfo  {journal}
  {Soft Matter}\ }\textbf {\bibinfo {volume} {6}},\ \bibinfo {pages} {5327}
  (\bibinfo {year} {2010})}\BibitemShut {NoStop}%
\bibitem [{\citenamefont {Kaz}\ \emph {et~al.}(2012)\citenamefont {Kaz},
  \citenamefont {McGorty}, \citenamefont {Mani}, \citenamefont {Brenner},\ and\
  \citenamefont {Manoharan}}]{Manoharan}%
  \BibitemOpen
  \bibfield  {author} {\bibinfo {author} {\bibfnamefont {D.~M.}\ \bibnamefont
  {Kaz}}, \bibinfo {author} {\bibfnamefont {R.}~\bibnamefont {McGorty}},
  \bibinfo {author} {\bibfnamefont {M.}~\bibnamefont {Mani}}, \bibinfo {author}
  {\bibfnamefont {M.~P.}\ \bibnamefont {Brenner}}, \ and\ \bibinfo {author}
  {\bibfnamefont {V.~N.}\ \bibnamefont {Manoharan}},\ }\href
  {http://dx.doi.org/10.1038/nmat3190} {\bibfield  {journal} {\bibinfo
  {journal} {Nat. Mater}\ }\textbf {\bibinfo {volume} {11}},\ \bibinfo {pages}
  {138} (\bibinfo {year} {2012})}\BibitemShut {NoStop}%
\bibitem [{\citenamefont {Colosqui}\ \emph {et~al.}(2013)\citenamefont
  {Colosqui}, \citenamefont {Morris},\ and\ \citenamefont {Koplik}}]{Carlos}%
  \BibitemOpen
  \bibfield  {author} {\bibinfo {author} {\bibfnamefont {C.~E.}\ \bibnamefont
  {Colosqui}}, \bibinfo {author} {\bibfnamefont {J.~F.}\ \bibnamefont
  {Morris}}, \ and\ \bibinfo {author} {\bibfnamefont {J.}~\bibnamefont
  {Koplik}},\ }\href {\doibase 10.1103/PhysRevLett.111.028302} {\bibfield
  {journal} {\bibinfo  {journal} {Phys. Rev. Lett.}\ }\textbf {\bibinfo
  {volume} {111}},\ \bibinfo {pages} {028302} (\bibinfo {year}
  {2013})}\BibitemShut {NoStop}%
\bibitem [{\citenamefont {Razavi}\ \emph {et~al.}(2014)\citenamefont {Razavi},
  \citenamefont {Kretzschmar}, \citenamefont {Koplik},\ and\ \citenamefont
  {Colosqui}}]{Sepideh}%
  \BibitemOpen
  \bibfield  {author} {\bibinfo {author} {\bibfnamefont {S.}~\bibnamefont
  {Razavi}}, \bibinfo {author} {\bibfnamefont {I.}~\bibnamefont {Kretzschmar}},
  \bibinfo {author} {\bibfnamefont {J.}~\bibnamefont {Koplik}}, \ and\ \bibinfo
  {author} {\bibfnamefont {C.~E.}\ \bibnamefont {Colosqui}},\ }\href {\doibase
  10.1063/1.4849135} {\bibfield  {journal} {\bibinfo  {journal} {J. Chem.
  Phys.}\ }\textbf {\bibinfo {volume} {140}} (\bibinfo {year} {2014}) }\BibitemShut {NoStop}%
\bibitem [{\citenamefont {Lamb}(1945)}]{Lamb}%
  \BibitemOpen
  \bibfield  {author} {\bibinfo {author} {\bibfnamefont {H.}~\bibnamefont
  {Lamb}},\ }\href@noop {} {\emph {\bibinfo {title} {Hydrodynamics}}}\
  (\bibinfo  {publisher} {Dover Publications},\ \bibinfo {year}
  {1945})\BibitemShut {NoStop}%
\bibitem [{\citenamefont {Brenner}(1962)}]{Brenner}%
  \BibitemOpen
  \bibfield  {author} {\bibinfo {author} {\bibfnamefont {H.}~\bibnamefont
  {Brenner}},\ }\href {\doibase 10.1017/S0022112062000026} {\bibfield
  {journal} {\bibinfo  {journal} {J. Fluid Mechanics}\ }\textbf {\bibinfo
  {volume} {12}},\ \bibinfo {pages} {35} (\bibinfo {year} {1962})}\BibitemShut
  {NoStop}%
\end{thebibliography}
\end{document}